\documentclass[aps,prd,twocolumn,twoside,superscriptaddress,nofootinbib,preprintnumbers]{revtex4-2}

\usepackage[bookmarks=false]{hyperref}
\usepackage{natbib}
\usepackage{graphicx}
\usepackage{amsmath}
\usepackage{amssymb}
\usepackage{siunitx}
\usepackage{mathrsfs}
\usepackage{xspace}
\usepackage{xcolor}
\usepackage[normalem]{ulem}
\usepackage[abs]{overpic}
\usepackage{comment}
\usepackage{dsfont}
\usepackage{bm}
\usepackage{subfigure}
\usepackage{cleveref}
\usepackage{braket}

\definecolor{darkblue}{cmyk}{1,0.4,0,0.3}
\definecolor{violet}{cmyk}{0,1,0,0.2}
\hypersetup{colorlinks, bookmarksnumbered, citecolor=darkblue, linkcolor=darkblue, pdfstartview=FitH, urlcolor=darkblue, linktocpage}

\def\d{\mathrm{d}}
\def\H{\bm{\mathcal{H}}}
\def\S{\bm{\mathcal{S}}}

\usepackage{lipsum}
\begin{document}
	
	\title{Nested-GPT for variable-multiplicity parton showers: \\ A case study in the resummation of non-global logarithms}
	
	\author{Wanchen Li}
	\email{wanchenli@fudan.edu.cn}
	\affiliation{Department of Physics and Center for Field Theory and Particle Physics, Fudan University, Shanghai, 200433, China}

	\author{Ding Yu Shao}
	\email{dyshao@fudan.edu.cn}
	\affiliation{Department of Physics and Center for Field Theory and Particle Physics, Fudan University, Shanghai, 200433, China}
	\affiliation{Key Laboratory of Nuclear Physics and Ion-beam Application (MOE), Fudan University, Shanghai, 200433, China}
	\affiliation{Shanghai Research Center for Theoretical Nuclear Physics, NSFC and Fudan University, Shanghai 200438, China}
	\affiliation{Center for High Energy Physics, Peking University, Beijing 100871, China}
	\affiliation{Southern Center for Nuclear-Science Theory (SCNT), Institute of Modern Physics, Chinese Academy of Sciences, Huizhou 516000, Guangdong Province, China}
	
	\author{Hao-Zhe Shi}
	\email{haozheshi869@gmail.com}
	\affiliation{Department of Physics and Center for Field Theory and Particle Physics, Fudan University, Shanghai, 200433, China}
	
	\author{Yu-Xuan Sun}
	\email{yxsun24@m.fudan.edu.cn}
	\affiliation{Department of Physics and Center for Field Theory and Particle Physics, Fudan University, Shanghai, 200433, China}
	
	%%%%%%%%%%%%%%%%%%%%%%%%%%%%%%%%%%%%%%%%%%%%%%%%%%%%%%%%%%%%%%%%%%%%%%%%%%%%%%%%%
	\begin{abstract}
		We introduce Nested-GPT, a hierarchical autoregressive Transformer architecture for simulating the variable-multiplicity parton-shower histories. As a controlled benchmark,  we study the leading-logarithmic resummation of non-global logarithms in the large-$N_c$ limit, utilizing a stochastic Monte Carlo dipole shower to generate reference training data. We systematically evaluate Nested-GPT against a Transformer flow-matching baseline. The flow-matching framework successfully parameterizes the joint distribution of emission kinematics at fixed multiplicity. Its phase-space representation, however, requires the final number of emissions to be specified externally rather than generated dynamically. Conversely, Nested-GPT follows the ordered Markovian branching structure, predicting emissions sequentially and dynamically evaluating a learned sequence-termination condition. We benchmark both approaches using gap fraction observables under two complementary training regimes: direct training on vetoed histories and inclusive training followed by an analysis-level veto. The resulting generated samples agree with the reference shower within statistical uncertainties for the observables considered. These results establish Nested-GPT as a physically consistent autoregressive surrogate for variable-multiplicity shower generator and motivate extensions to subleading-logarithmic resummation and finite-$N_c$ color evolution.
	\end{abstract}
	%%%%%%%%%%%%%%%%%%%%%%%%%%%%%%%%%%%%%%%%%%%%%%%%%%%%%%%%%%%%%%%%%%%%%%%%%%%%%%%%%
	\maketitle

	%%%%%%%%%%%%%%%%%%%%%%%%%%%%%%%%%%%%%%%%%%%%%%%%%%%%%%%%%%%%%%%%%%%%%%%%%%%%%%%%%
	\section{Introduction}
	
	Artificial intelligence and machine learning have emerged as powerful tools across various subfields of high-energy physics, from amplitude calculations and phase-space integration to detector simulation and statistical inference. 
	These tools encompass several architectural families---supervised classifiers, generative adversarial networks, normalizing flows, diffusion models, and autoregressive transformers---and are now deployed throughout the Large Hadron Collider (LHC) theory and experiment pipeline~\cite{Butter:2022rso, Ubiali:2026, Cai:2026ths}. 
	Generative models, in particular, now serve as surrogates for a wide range of event-generation tasks: GANs for calorimeter and jet-image simulation~\cite{Goodfellow:2014upx, Paganini:2017hrr, deOliveira:2017pjk} and full event-level generation~\cite{Butter:2019cae, Butter:2019eyo, Butter:2020qhk}; normalizing flows for variable-multiplicity event generation and phase-space integration~\cite{Papamakarios:2021flows, Butter:2021csz, Heimel:2022wyj, Gao:2020zvv, Bothmann:2020ywa}; and diffusion and flow-matching models for collider event simulation~\cite{Ho:2020ddpm, Song:2021sde, Lipman:2023flowmatching, Albergo:2023interpolants, Mikuni:2022xry, Buhmann:2023epicfm, Butter:2023fov}. 
	A pressing application area for these architectures is the simulation of strong-interaction processes at colliders, where they either substantially accelerate numerical methods or reshape the computational workflow substantially~\cite{Plehn:2022059, hepmllivingreview}. 
	
	Quantum chromodynamics (QCD) describes the strong interaction and provides the theoretical foundation for collider phenomenology~\cite{Gross_2023}. 
	It describes the partonic structure of the initial state~\cite{Martin:2009iq} and dictates the evolution of the hard scattering into hadronic final states. QCD radiation and QCD-induced final states constitute the dominant background for new physics searches at the LHC~\cite{Kogler:2018hem}.
	Monte Carlo event generators simulate this evolution primarily through the parton shower, providing a probabilistic description of sequential QCD radiation~\cite{Buckley:2011ms}. Starting from a hard partonic configuration, showers generate successive emissions typically ordered in an evolution variable, resumming the logarithmically enhanced soft and collinear radiation that dominates exclusive observables.
	This evolution is governed by the renormalization group equations and the associated splitting kernels; practically, it is realized as a Markov chain of $1 \to 2$ branchings, with the non-branching probability encoded in the Sudakov form factor. 
	Machine learning has augmented this domain by training neural networks to reweight predictions for uncertainty exploration~\cite{Bothmann:2018}, utilizing autoencoders to mirror the renormalization group~\cite{Monk:2018}, and employing transformers for the autoregressive generation of $\text{Z}$+jets events~\cite{Butter:2024zbd, Butter:2023fov}. 
	Beyond the shower itself, machine learning is transforming the broader event-generation pipeline: normalizing flows and neural importance sampling now accelerate phase-space integration and event unweighting~\cite{Bothmann:2020ywa, Gao:2020zvv, Heimel:2022wyj, danziger_accelerating_2022}, with recent extensions to NLO calculations~\cite{crescenzo_madnis_2026} and dedicated GPU-native frameworks like MadSpace~\cite{heimel_madspace_2026}, while machine learning surrogates speed up matrix-element reweighting from leading-color to full-color accuracy~\cite{villadamigo_fastcolor_2025}.

	While standard general-purpose showers in \textsc{Pythia}~\cite{Bierlich:2022pfr}, \textsc{Sherpa}~\cite{Sherpa:2024mfk}, and \textsc{Herwig}~\cite{Bellm:2015jjp} provide successful leading-logarithmic (LL) descriptions, achieving systematic logarithmic control beyond LL remains an active frontier~\cite{Nagy:2020rmk, Forshaw:2020wrq}. 
	The PanScales collaboration has demonstrated full next-to-leading-logarithmic (NLL) and next-to-next-to-leading-logarithmic (NNLL) accuracy for global event shapes~\cite{vanBeekveld:2024wws, PanScales:2025}.
	However, non-global logarithms (NGLs) arise in observables sensitive to restricted phase-space regions, such as jet vetoes and gaps between jets. These observables encode correlated wide-angle emissions that do not trivially exponentiate \cite{Dasgupta:2001sh}. At LL accuracy in the large-$N_c$ limit, NGL resummation is governed by the non-linear Banfi--Marchesini--Smye (BMS) equation~\cite{Banfi:2002hw}, physically interpreted as a dipole shower in the rapidity-azimuth plane~\cite{Dasgupta:2001sh}. 
	This non-linear NGL evolution connects to the Balitsky--Kovchegov (BK) equations~\cite{Balitsky:1998ya, Kovchegov:1999ua, Kovchegov:1999yj}, which govern small-$x$ gluon saturation, via conformal duality~\cite{Caron-Huot:2015bja, Brunello:2025rhh}. 
	Extending this framework to full color or capturing Glauber-gluon exchanges in the hadronic environment of the LHC introduces severe computational bottlenecks~\cite{Forshaw:2006fk, Hatta:2013iba, DeAngelis:2020rvq}. 
	Achieving higher-logarithmic accuracy for non-global observables~\cite{Banfi:2021owj, Banfi:2021xzn, Becher:2021urs, Becher:2023vrh, FerrarioRavasio:2023kyg} necessitates algorithms that substantially exceed the capabilities of current public codes.

	These bottlenecks present an underexplored intersection between high-precision QCD and machine learning. 
	Despite the emergence of autoregressive Transformers in jet constituent generation~\cite{Butter:2024zbd, Butter:2023fov}, existing architectures like PC-JeDi~\cite{Leigh:2023toe} or JetGPT~\cite{Butter:2023fov, Butter:2024zbd} are not designed to explicitly encode the Markovian structure of parton showers or the Sudakov-governed termination condition intrinsic to physical showers. One solution to this problem is provided by the DGLAP-based parton shower model~\cite{Lai:2020byl}, which explicitly encodes the recursive $1\rightarrow2$ splitting process and learns a Sudakov-like termination condition. However, due to its inherently sequential generation architecture, this approach cannot be efficiently scaled to multi-jet events.
	For NGL resummation, the large-$N_c$ RG evolution naturally produces ordered sequences of emissions in the shower-time variable~\cite{Balsiger:2018ezi, Balsiger:2020ogy}. 
	Generating high-statistics samples of these discrete shower histories beyond LL accuracy is computationally prohibitive due to the complex non-linear structure of the NGL evolution~\cite{Banfi:2021owj, Banfi:2021xzn, Becher:2021urs, Becher:2023vrh}. 
	
	To address this, we introduce Nested-GPT, a hierarchical autoregressive Transformer designed to emulate NGL resummation. 
	By evaluating an explicit stop/continue decision after every emission step, Nested-GPT learns the Sudakov-driven termination pattern of the reference shower while preserving the shower-time ordering of the generated sequence. 
	Concurrently, the Transformer's attention mechanism captures the complex, non-local dipole correlations characteristic of the non-linear NGL evolution, bypassing the computational limits of standard Monte Carlo integration. 
	We evaluate this architecture alongside a fixed-multiplicity flow-matching baseline, which, while useful, lacks the intrinsic termination condition necessary for a fully automated generator. 
	By applying phase-space restrictions to the generated ensemble, we obtain neural approximations to the corresponding LL large-\(N_c\) dipole-shower predictions.
	This formulation provides a computationally viable surrogate capable of generating high-statistics shower histories, establishing a promising framework to evaluate complex non-global theoretical predictions against LHC data in future.
	
	The remainder of the paper is organized as follows. In Sec.~\ref{sec:formalism_gap} we briefly review the renormalization group (RG) structure underlying NGL resummation and the BMS equation, summarizing the dipole-shower Monte Carlo solution at LL. In Sec.~\ref{sec:models} we define the event representation and introduce the generative models. Results are presented in Sec.~\ref{sec:results}.
	
	\section{Parton shower approach to NGL resummation}\label{sec:formalism_gap}
	
	NGLs emerge when an observable constrains radiation in only a restricted region of phase space. A seminal example is the gap fraction, where emissions in the gap region are restricted below a soft scale $Q_0$. Because emissions outside the gap can radiate secondary soft partons into the gap region, the probability of surviving the veto cannot be described by a simple independent-emission Sudakov factor. Instead, correlated soft radiation induces a non-linear evolution driven by connected color flows between the vetoed and inclusive regions.
	
	In this section, we first review the BMS equation, which established the standard LL large-$N_c$ framework and admits a transparent physical interpretation in terms of dipole branching. We then introduce the RG evolution formalism from an effective field theory perspective. This formalism systematically generalizes the BMS framework and translates directly into a dipole-shower Monte Carlo algorithm at LL in the large-$N_c$ limit \cite{Becher:2015hka, Becher:2016mmh, Balsiger:2018ezi, Balsiger:2020ogy}.
	
	Consider a back-to-back \(q\bar q\) dipole produced in \(e^+e^-\) annihilation at a hard center-of-mass scale \(Q\). The gap region is geometrically defined by $|y|<y_{\max}$, where the rapidity $y$ of a soft emission with momentum $k^\mu=(k^0,\vec{k})$ is evaluated with respect to the thrust axis $\vec{n}$ as
	\begin{align}
		y
		=
		\frac12\ln\frac{k^0+\vec n\cdot\vec k}{k^0-\vec n\cdot\vec k}\,.
		\label{eq:rapidity_def}
	\end{align}
	The gap fraction $G(Q,Q_0)$ is defined as the ratio of the vetoed cross section $\sigma_{\rm veto}(Q,Q_0)$—where energy deposited into the veto region is strictly less than $Q_0$—to the total cross section $\sigma_{\rm tot}(Q)$,
	\begin{align}
		G(Q,Q_0)
		\equiv
		\frac{\sigma_{\rm veto}(Q,Q_0)}{\sigma_{\rm tot}(Q)}\,.
		\label{eq:gap_fraction}
	\end{align}
	
	\subsection{The BMS equation}\label{sec:bms_recap}
	
	At LL accuracy in the large-$N_c$ limit, the resummation of NGLs is governed by the BMS equation \cite{Banfi:2002hw}. This non-linear integro-differential equation dictates the scale evolution of the gap fraction $G_{ab}(Q)$, initiated by a primary dipole $(ab)$ with respect to the scale $Q$
	\begin{align}
		Q\,\frac{\partial}{\partial Q}\,G_{ab}(Q)
		=&\;
		\int \frac{\d^2\Omega_k}{4\pi}\,\bar\alpha_s\,
		W_{ab}(k) \nonumber \\
		&\hspace{-1cm} \times \Bigl[
		u(k)\,G_{ak}(Q)\,G_{kb}(Q)
		-G_{ab}(Q)
		\Bigr],
		\label{eq:BMS_basic}
	\end{align}
	where the effective coupling is $\bar{\alpha}_s \equiv \alpha_s N_c / \pi$. Here, $\d^2\Omega_k$ denotes the angular integration measure for the radiated soft gluon $k$, and the eikonal dipole radiator is given by $W_{ab}(k) = (n_a \cdot n_b) / [(n_a \cdot n_k)(n_k \cdot n_b)]$. The measurement function $u(k)$ enforces the phase-space geometry, evaluating to zero if the emission falls within the restricted gap region and unity otherwise
	\begin{align}
		u(k) =
		\begin{cases}
			1\,, & k \ \text{outside the gap},\\
			0\,, & k \ \text{inside the gap}.
		\end{cases}
	\end{align}
	
	While \cref{eq:BMS_basic} governs the deterministic evolution of an inclusive probability, its physical structure naturally admits a stochastic, Markovian interpretation. The real-emission term $u(k)G_{ak}G_{kb}$ encapsulates the branching of the parent dipole $(ab)$ into two independent daughter dipoles, $(ak)$ and $(kb)$, a factorization strictly valid only in the large-$N_c$ limit. Conversely, the virtual term $-G_{ab}$ accounts for the exact non-branching probability. This iterative dipole proliferation, wherein each successive unvetoed emission acts as a novel radiator, constitutes the dynamical engine of the non-linear evolution and the fundamental source of non-global correlations.

	\subsection{RG evolution of NGLs}
	\label{subsec:RG_NGL_gap}
	
	The factorization theorem, formulated within the framework of SCET, provides a robust and comprehensive foundation for describing non-global observables. For an exclusive $k$-jet vetoed cross section at an $e^+e^-$ collider, the observable rigorously factorizes into hard functions $\H$, which encapsulate energetic radiation in the unconstrained phase space, and soft functions $\S$, which describe emissions originating from Wilson lines along the hard-parton directions,
	\begin{equation}
		\d\sigma(Q,Q_0)
		=
		\sum_{m=k}^{\infty}
		\left\langle
		\H_m(\{n\},Q,\mu)\otimes \S_m(\{n\},Q_0,\mu)
		\right\rangle ,
		\label{eq:fact_ngl_gap}
	\end{equation}
	Here, \(\{n\} = \{n_1,\ldots,n_m\}\) denotes the hard parton directions, \(\otimes\) indicates angular integration over these directions, $\mu$ is the renormalization scale, and \(\langle\ldots\rangle\) denotes the color trace. A defining hallmark of non-global observables is the sensitivity of soft radiation to the complete topological configuration of hard partons in the event. Due to this sensitivity, operators characterized by different numbers of Wilson lines mix under renormalization. Consequently, the associated RG evolution intrinsically assumes a matrix-valued structure in the $m$-parton multiplicity space. While we direct the reader to Refs.~\cite{Becher:2015hka, Becher:2016mmh, Balsiger:2018ezi} (see also \cite{Larkoski:2015zka, Caron-Huot:2015bja}) for the rigorous theoretical definitions of $\H$ and $\S$, our primary focus here lies in the structural dynamics of this evolution.
	
	Imposing the requirement that the physical vetoed cross section in \cref{eq:fact_ngl_gap} remain invariant under variations of the renormalization scale, one finds that the scale dependence of the hard functions strictly dictates the corresponding RG evolution of the soft sector. The hard functions satisfy the matrix RG equation
	\begin{equation}
		\frac{\d}{\d\ln\mu} \H_m(\{n\},Q,\mu)
		=
		- \sum_{l=k}^{m}
		\H_l(\{n\},Q,\mu)\,\bm \Gamma^{H}_{lm}(\{n\},Q,\mu),
		\label{eq:hard_RG}
	\end{equation}
	yielding a formal solution expressed via the path-ordered evolution operator
	\begin{equation}
		\bm U(\{n\},\mu_s,\mu_h)
		=
		\mathbf{P}\exp\!\left[
		\int_{\mu_s}^{\mu_h}\frac{\d\mu}{\mu}\,
		\bm \Gamma^H(\{n\},\mu)
		\right].
		\label{eq:U_pathordered}
	\end{equation}
	By systematically evolving the hard functions from their characteristic hard scale \(\mu_h\sim Q\) down to the soft scale \(\mu_s\sim Q_0\), the fully resummed cross section is expressed as
	\begin{align}
		\d\sigma(Q,Q_0)
		&=
		\sum_{l=k}^{\infty}\sum_{m\ge l}
		\langle
		\H_l(\{n\},Q,\mu_h) \nonumber \\
		&\hspace{-0.5cm}\otimes
		\bm U_{lm}(\{n\},\mu_s,\mu_h)\,\hat\otimes\,\S_m(\{n\},Q_0,\mu_s)
		\rangle ,
		\label{eq:resummed_cross_section}
	\end{align}
	where \(\hat\otimes\) encapsulates the angular integrations over the phase space of the ($m-l$) unresolved emissions. Within this formalism, the resummation of NGLs is mapped directly onto the exponentiation of an infinite-dimensional anomalous-dimension matrix.
	
	At LL accuracy, this matrix architecture simplifies dramatically. The soft functions trivially reduce to the identity matrix in color space, \(\S_m=\bm 1\), and higher-multiplicity hard functions evaluated at $\mu_h$ become suppressed. Thus, the LL cross section assumes the compact form
	\begin{align}
		\d\sigma^{\rm LL}&(Q,Q_0)
		= \\
		&\sum_{m=k}^{\infty}
		\left\langle
		\bm \H_k(\{n\},Q,\mu_h)\otimes
		\bm U_{km}(\{n\},\mu_s,\mu_h)\,\hat\otimes\,\bm 1
		\right\rangle .\notag
		\label{eq:LL_cross_section}
	\end{align}
	Furthermore, at LL accuracy, the evolution trajectory is governed entirely by the one-loop anomalous dimension. It is therefore highly convenient to reparameterize the exponent of the evolution matrix as
	\begin{equation}
		\int_{\mu_s}^{\mu_h}\frac{\d\mu}{\mu}\bm \Gamma^H_{lm}
		=
		\int_{\alpha_s(\mu_h)}^{\alpha_s(\mu_s)} \frac{\d\alpha}{\beta(\alpha)}
		\frac{\alpha}{4\pi}\bm \Gamma^{(1)}_{lm}
		=\frac{1}{2\beta_0}
		\ln\frac{\alpha_s(\mu_s)}{\alpha_s(\mu_h)}\bm \Gamma^{(1)}_{lm}.
		\label{eq:tGamma}
	\end{equation}
	This substitution naturally introduces the dimensionless evolution time $t$, defined as
	\begin{equation}
		t=\frac{1}{2\beta_0}
		\ln\frac{\alpha_s(\mu_s)}{\alpha_s(\mu_h)}
		=
		\frac{\alpha_s}{4\pi}\ln\frac{\mu_h}{\mu_s}
		+\mathcal{O}(\alpha_s^2),
		\label{eq:evolution_time}
	\end{equation}
	which provides a logarithmic measure of the separation between the hard and soft scales. Physically, \(t=0\) represents the unevolved system (\(\mu_s=\mu_h\)), with increasing values of \(t\) corresponding to an RG evolution spanning a progressively wider scale hierarchy.
	
	Structurally, the one-loop anomalous-dimension matrix adopts an upper-triangular form,
	\begin{equation}
		\bm \Gamma^{(1)}=
		\begin{pmatrix}
			\bm V_k & \bm R_k & 0 & 0 & \cdots \\
			0 & \bm V_{k+1} & \bm R_{k+1} & 0 & \cdots \\
			0 & 0 & \bm V_{k+2} & \bm R_{k+2} & \cdots \\
			0 & 0 & 0 & \bm V_{k+3} & \cdots \\
			\vdots & \vdots & \vdots & \vdots & \ddots
		\end{pmatrix},
		\label{eq:Gamma_triangular}
	\end{equation}
	where \(\bm V_m\) and \(\bm R_m\) denote the virtual and real-emission kernels, respectively. In full QCD, evaluating the action of these kernels entails highly non-trivial kinematics coupled with intricate color algebra. Because the relevant color space dimension grows exponentially with the parton multiplicity \(m\), exact numerical evaluation rapidly becomes computationally prohibitive.
	
	Crucially, this computational bottleneck is broken in the large-$N_c$ limit. Under this approximation, the action of the real emission kernel $R_m$ on source partons ($i$, $i+1$) restricts the color flow such that the new \((m+1)\)-th emission couples exclusively to adjacent partons. As a result, the multi-parton ensemble dynamically fragments into an independent sequence of disconnected color dipoles. The exact evolution kernels thus simplify to
	\begin{align}
		\bm V_m
		= &
		-4N_c \, \bm 1 \sum_{i=k}^{m-1} \int \frac{\d\Omega(n_{m+1})}{4\pi}\,W^{m+1}_{i,i+1}.
		\label{eq:Vm_largeNc} \\
		\bm R_m
		=&\; 4N_c \, \bm 1 \sum_{i=k}^{m-1} W^{m+1}_{i,i+1}\Theta_{\rm in}(n_{m+1}).
		\label{eq:Rm_largeNc}
	\end{align}
	
	\subsection{Parton-shower solution of the RG equation}
	\label{subsec:parton_shower_solution}
	
	\begin{table*}[t]
		\caption{Monte Carlo realization of the LL NGL RG evolution in the large-\(N_c\) limit,
			adapted from Ref.~\cite{Balsiger:2018ezi}. For simplicity, we suppress the detailed
			weighting factors associated with the generated emission configurations.}
		\label{tab:MC_algorithm}
		\vspace{0.1cm}
		\begin{tabular}{c c}
			\hline\hline
			Step & Description \\
			\hline
			1 &
			\parbox[t]{0.92\textwidth}{\raggedright
				Initialize the shower at \(t=0\) with the Born event \(E=\{n_1,n_2\}\) and compute the total virtual rate \(V_E\) from \cref{eq:VE_final}.
				\vspace*{0.01cm}
			}
			\\
			
			2 &
			\parbox[t]{0.92\textwidth}{\raggedright
				Generate a random time step \(\Delta t\) from the exponential distribution
				$
				P_E(\Delta t)=V_E\,e^{-V_E\Delta t}.
				$.
				\vspace*{0.25cm}
			}
			\\
			
			3 &
			\parbox[t]{0.92\textwidth}{\raggedright
				Choose one dipole \((ij)\in E\) within the entire dipole configuration with probability \(V_{ij}/V_E\). Generate a new particle emission in a random direction \(n_k\), and update the event record.
				\vspace*{0.25cm}
			}
			\\ 
			
			4 &
			\parbox[t]{0.92\textwidth}{\raggedright
				If \(n_k\) lies in the veto region, terminate the event and restart from Step~1. Otherwise update the ordered list by inserting the new direction between the emitting legs,
				$E=\{n_1,\ldots,n_i,n_j,\ldots,n_2\}
				\to
				E'=\{n_1,\ldots,n_i,n_k,n_j,\ldots,n_2\}
				$, set \(t\to t+\Delta t\), and return to Step~2.
			}
			
			\\[0.5cm]
			\hline\hline
		\end{tabular}
	\end{table*}

	A numerical parton-shower framework for solving the LL RG equation is naturally formulated by exploiting the upper-triangular structure of the one-loop anomalous-dimension matrix. Rewriting the scale evolution in terms of the dimensionless evolution time \(t\), the underlying RG equation 
	\begin{equation}
		\frac{\d}{\d t} \H_m(t)
		=
		\H_m(t)\,\bm V_m + \H_{m-1}(t)\,\bm R_{m-1},
		\label{eq:RG_t_form}
	\end{equation}
	is recast into the integral form
	\begin{align}
		\H_m(t)
		=&\;
		\H_m(t_0)e^{(t-t_0) \bm V_m}
		\nonumber \\
		&+
		\int_{t_0}^{t}\d t'\,
		\H_{m-1}(t')\,\bm R_{m-1}\,e^{(t-t')\bm V_m}.
		\label{eq:integral_recursive}
	\end{align}
	This formulation manifestly exposes the recursive structure characteristic of a parton shower. Specifically, an $m$-parton configuration either evolves continuously over time without radiation (captured by the first line of \cref{eq:integral_recursive}), or it is dynamically generated via a single real emission branching from a lower-multiplicity state (the second line). Consequently, starting from the initial $m=k$ Born state, the successive hard functions at higher multiplicities $m=k+1, k+2, \dots$ are generated iteratively,
	\begin{align}
		\H_k(t) &= \H_k(0)e^{t \bm V_k}, \nonumber\\
		\H_{k+1}(t) &= \int_0^t \d t' \, \H_k(t') \bm R_k e^{(t-t') \bm V_{k+1}}, \nonumber\\
		\H_{k+2}(t) &= \int_0^t \d t' \, \H_{k+1}(t') \bm R_{k+1} e^{(t-t') \bm V_{k+2}}, \nonumber \\
		\H_{k+3}(t) &= \dots \,.
		\label{eq:iterative_Hm}
	\end{align}

	Driven by this underlying recursive architecture, the expansion defined in \cref{eq:integral_recursive,eq:iterative_Hm} is perfectly suited for a direct Monte Carlo realization. In the large-\(N_c\) limit, an arbitrary multi-parton configuration can be uniquely parameterized by an ordered sequence of Wilson-line directions,
	\begin{equation}
		E=\{n_1,n_{i_1},n_{i_2},\ldots,n_{i_{m-2}},n_2\},
	\end{equation}
	where adjacent elements in the sequence specify the active color dipoles of the event. For simplicity, we select a back-to-back dijet topology ($k=2$) as the Born configuration, featuring initial parton directions $n_1$ and $n_2$. A fundamental theoretical quantity governing the showering algorithm is the total virtual decay rate associated with a given topological state \(E\),
	\begin{align}
		V_E
		&= V_{1i_1}+V_{i_1i_2}+\cdots+V_{i_{m-2}2},
		\label{eq:VE_final}
		\\
		V_{ij}
		&= \int \frac{\d\Omega(n_l)}{4\pi}\,R_{ij}^l,
		\label{eq:Vij_final}
		\\
		R_{ij}^l
		&= 4N_c\,W_{ij}^l\,
		\theta(n_l\!\cdot n_i-\lambda^2)\,
		\theta(n_l\!\cdot n_j-\lambda^2).
		\label{eq:Rijl_final}
	\end{align}
	Here, \(R_{ij}^l\) denotes the regulated real-emission matrix element, while \(V_{ij}\) represents the corresponding integrated virtual contribution. To explicitly regulate collinear divergences, we introduce a collinear cutoff parameter $\lambda$, contrasting with the unregulated real-emission kernel previously defined in \cref{eq:Rm_largeNc}.

	The core operations of the resulting parton-shower algorithm are summarized in \cref{tab:MC_algorithm}. The physical interpretation of this procedure is highly intuitive. The exponential Sudakov factor \(\exp(-V_E\Delta t)\) in \cref{eq:integral_recursive} dictates the standard non-emission probability, characterizing the likelihood that configuration \(E\) survives over a shower-time interval \(\Delta t\) without undergoing a branching. Upon a branching event, the algorithm stochastically selects an active emitting dipole with probability \(V_{ij}/V_E\), and subsequently samples the direction \(n_k\) of the new emission from the normalized kernel \(R_{ij}^k/V_{ij}\). Crucially, if the generated emission falls inside the experimentally vetoed phase-space region, the event violates the observable constraints and is immediately terminated. Otherwise, the newly generated parton is inserted between the two constituent legs of the emitting dipole. This operation effectively replaces the parent dipole \((ij)\) with two daughter dipoles \((ik)\) and \((kj)\), allowing the evolution to iteratively proceed from this updated topology.
	
	Ultimately, the LL resummation of NGLs is elegantly mapped onto a Markovian branching process evolving continuously in the shower time \(t\). Within this paradigm, the simulated parton shower serves as the exact stochastic solution to the RG evolution equation governing the non-global observable. We refer the reader to Ref.~\cite{Balsiger:2018ezi} for comprehensive technical details regarding the specific implementation of showering weights.
	
	\section{Models}
	\label{sec:models}
	
	In this section, we detail the two generative models trained to simulate dipole-shower histories: a flow-matching baseline (\cref{sec:flow-matching}) and a Nested-GPT architecture (\cref{sec:nested-gpt}).

	\subsection{Flow-matching baseline}
	\label{sec:flow-matching}
	
	For the flow-matching model, we represent each dipole-shower history as a single multi-particle event $E=\{(t_j,\mathbf{p}_j)\}_{j=1}^{N}$, ordered by monotonically increasing shower time $t_j$, where $\mathbf{p}_j$ denotes the three momentum $(p_{x,j},p_{y,j},p_{z,j})$. We enforce a maximum multiplicity of $N_{\max}=100$ emissions per event, zero-padding shorter sequences and utilizing a boolean mask $m_j\in\{0,1\}$ to distinguish physical emissions ($m_j=1$ for $j\le N$) from padded tokens. The dataset statistics are summarized in Table~\ref{tab:dataset-summary}.
	
	We introduce a continuous-time generative model based on flow matching to map the distribution of events within a continuous latent feature space \cite{Lipman:2023flowmatching, Albergo:2023interpolants}. To construct a representation suitable for regression, each emission $j$ is characterized by a four-dimensional feature vector
	\begin{equation}
		\mathbf{x}_j = \bigl(\log(\Delta t_j + \epsilon),\, p_{x,j},\, p_{y,j},\, p_{z,j}\bigr)\in\mathbb{R}^4,
		\label{eq:fm-features}
	\end{equation}
	where $\Delta t_j = t_j - t_{j-1}$ with $t_0 \equiv 0$ and $\epsilon=10^{-8}$ is a regularization constant for numerical stability. The three momentum components provide a smooth coordinate representation suitable for the flow-matching framework.
	
	\begin{table}[t]
		\caption{Dataset summary for the dipole-shower events used in this work. Counts are reported
			after truncating each event to at most $N_{\max}$ emissions.}
		\label{tab:dataset-summary}
		\begin{ruledtabular}
			\begin{tabular}{l c}
				Quantity & Value \\
				\hline
				Number of events & \num{970975} \\
				Number of emissions (total) & \num{7436741} \\
				Maximum multiplicity $N_{\max}$ & 100 \\
				Mean multiplicity $\langle N\rangle$ & \num{7.6590} \\
				Train/val split & $90\%/10\%$ \\
			\end{tabular}
		\end{ruledtabular}
	\end{table}
	
	The model architecture is defined by a time-dependent velocity field $v_\theta(\mathbf{x},\tau)$, which deterministically transports a tractable Gaussian base distribution to the target data distribution via the ordinary differential equation (ODE)
	\begin{equation}
		\frac{\d\mathbf{x}(\tau)}{\d\tau} = v_\theta\!\bigl(\mathbf{x}(\tau), \tau\bigr),
		\label{eq:fm-ode}
	\end{equation}
	with $\mathbf{x}(0)\sim\mathcal{N}(\mathbf{0},\mathbf{I})$ and $\tau\in[0,1]$. While this formulation is mathematically equivalent to a continuous normalizing flow \cite{Chen:2018neuralode}, we bypass likelihood optimization in favor of learning the transport velocity field directly through a regression objective. We use the standard linear interpolation path. For a given data sample $\mathbf{x}_1$, we draw latent noise $\mathbf{x}_0\sim\mathcal{N}(\mathbf{0},\mathbf{I})$ and an intermediate flow time $\tau\sim\mathcal{U}(0,1)$, defining the trajectory $\mathbf{x}_\tau = \tau\,\mathbf{x}_1 + (1-\tau)\,\mathbf{x}_0$ with target velocity $v^\star = \mathbf{x}_1 - \mathbf{x}_0$. The training objective is a masked mean-squared error (MSE) loss
	\begin{equation}
		\mathcal{L}_{\mathrm{FM}}
		= \mathbb{E}_{\mathbf{x}_1,\mathbf{x}_0,\tau}\!\left[
		\frac{1}{\sum_{j=1}^{N_{\max}} m_j}
		\sum_{j=1}^{N_{\max}} m_j
		\left\lVert v_\theta(\mathbf{x}_\tau,\tau)_j - v^\star_j \right\rVert_2^2
		\right].
		\label{eq:fm-loss}
	\end{equation}
	
	We parameterize $v_\theta$ using a Transformer encoder~\cite{Vaswani:2017attention}. Each particle feature $\mathbf{x}_j(\tau)$ is embedded into a hidden representation and augmented with sinusoidal positional encodings for both the particle index and the flow time $\tau$. Self-attention layers incorporate the padding mask to prevent information leakage from invalid (padded) tokens. A multi-layer perceptron (MLP) readout head predicts the per-particle velocity, taking the concatenation of the Transformer hidden state and the original feature vector as input. Our configuration employs $6$ Transformer layers with $8$ self-attention heads and a hidden dimension of $256$. Similar Transformer-based flow-matching architectures have been successfully applied to particle-cloud generation in collider physics~\cite{Buhmann:2023epicfm}.
	
	Sampling proceeds by first drawing an event multiplicity $N$ from the reference multiplicity distribution observed in the training data. We then sample $\mathbf{x}(0)\sim\mathcal{N}(\mathbf{0},\mathbf{I})$ with shape $N_{\max}\times4$, build the mask $m_j=1$ for $j\le N$ (and $m_j=0$ otherwise), and integrate the ODE in Eq.~\eqref{eq:fm-ode} from $\tau=0$ to $\tau=1$ using an explicit Euler solver with $K$ steps,
	\begin{align}
		\mathbf{x}^{(k+1)} &= \mathbf{x}^{(k)} + h\,\bigl(m\odot v_\theta(\mathbf{x}^{(k)},\tau_k)\bigr),
		\label{eq:fm-euler} \\
		h &= 1/K, \qquad \tau_k = k\,h.
	\end{align}
	where $\odot$ denotes element-wise multiplication broadcast over feature dimensions. Finally, we map the generated features back to physical observables. The inter-emission times are recovered as \(\Delta t_j=\exp(\hat{x}_{j,1})-\epsilon\), where \(\hat{x}_{j,1}\) is the first component of the generated feature vector. The absolute shower times are then obtained by a cumulative sum, transverse momentum, rapidity and azimuthal angle \((p_T,\eta,\phi)\) are computed from \((p_x,p_y,p_z)\) for evaluation and visualization. For the back-to-back Born configuration considered here, the pseudorapidity used in the generated representation coincides with the rapidity variable \(y\) defined in Eq.~\eqref{eq:rapidity_def} in the massless limit.
	
	For optimization we use AdamW (learning rate $10^{-4}$, weight decay $10^{-5}$), global batch size $512$ distributed across GPUs with DDP, and gradient clipping with maximum norm $1.0$ \cite{Loshchilov:2019adamw}. The learning rate is scheduled with a cosine decay and a linear warmup of $10\%$ of the total training steps. We report the masked validation loss from Eq.~\eqref{eq:fm-loss} each epoch and retain the checkpoint with the lowest validation loss. All splits are performed at the event level with a $90\%/10\%$ train/validation split.
	
	Because the multiplicity \(N\) is sampled and supplied to the model before ODE integration, the flow-matching baseline represents a conditional fixed-multiplicity generator rather than an unconditional variable-multiplicity shower. It therefore does not learn the emission and no-emission probabilities that determine shower termination. This limitation motivates the autoregressive construction described in Sec.~\ref{sec:nested-gpt}.
	
	\subsection{Nested-GPT}
	\label{sec:nested-gpt}

	To model variable-multiplicity shower histories end-to-end, we implement an autoregressive event generator inspired by GPT-style causal self-attention \cite{Radford:2018gpt, Radford:2019gpt2}. Similar approaches have recently been applied to event generation in collider physics \cite{Finke:2023ltl, Butter:2023fov}.
	
	Adopting the padded event representation detailed in Sec.~\ref{sec:flow-matching}, we define binary termination labels $s_i\in\{0,1\}$, where $s_i=1$ indicates a continuation signal ($i<N$) and $s_N=0$ denotes the sequence termination (stop) signal. Each emission $i$ is parameterized by a multi-dimensional feature vector
	\begin{equation}
		\mathbf{q}_i
		= \bigl(\log(\Delta t_i + \epsilon),\, \log(p_{T,i} + \epsilon),\, \eta_i,\, \phi_i \bigr),
		\label{eq:gpt-features}
	\end{equation}
	where $\phi_i\in[-\pi,\pi]$ and $\Delta t_i$ denotes the inter-emission time. To use a categorical likelihood, each feature dimension is discretized independently into $B$ bins using quantiles of the training data.\footnote{In the implementation we construct $B-1$ interior bin edges from quantiles in the range $[0.01,0.99]$ and add outer edges near the observed minimum and maximum to form $B$ bins. Values are clipped to the valid bin range at training and generation time.} Let $\{e_{f,k}\}_{k=0}^{B}$ be the bin edges and $c_{f,k}=\tfrac{1}{2}(e_{f,k}+e_{f,k+1})$ the corresponding bin centers for feature $f\in\{1,\dots,F\}$ with $F=4$. The discretized token is $z_{i,f}\in\{0,\dots,B-1\}$ and we denote the particle token vector by $\mathbf{z}_i=(z_{i,1},\dots,z_{i,F})$.
	
\begin{figure*}[t]
		\centering
		\includegraphics[width=\textwidth]{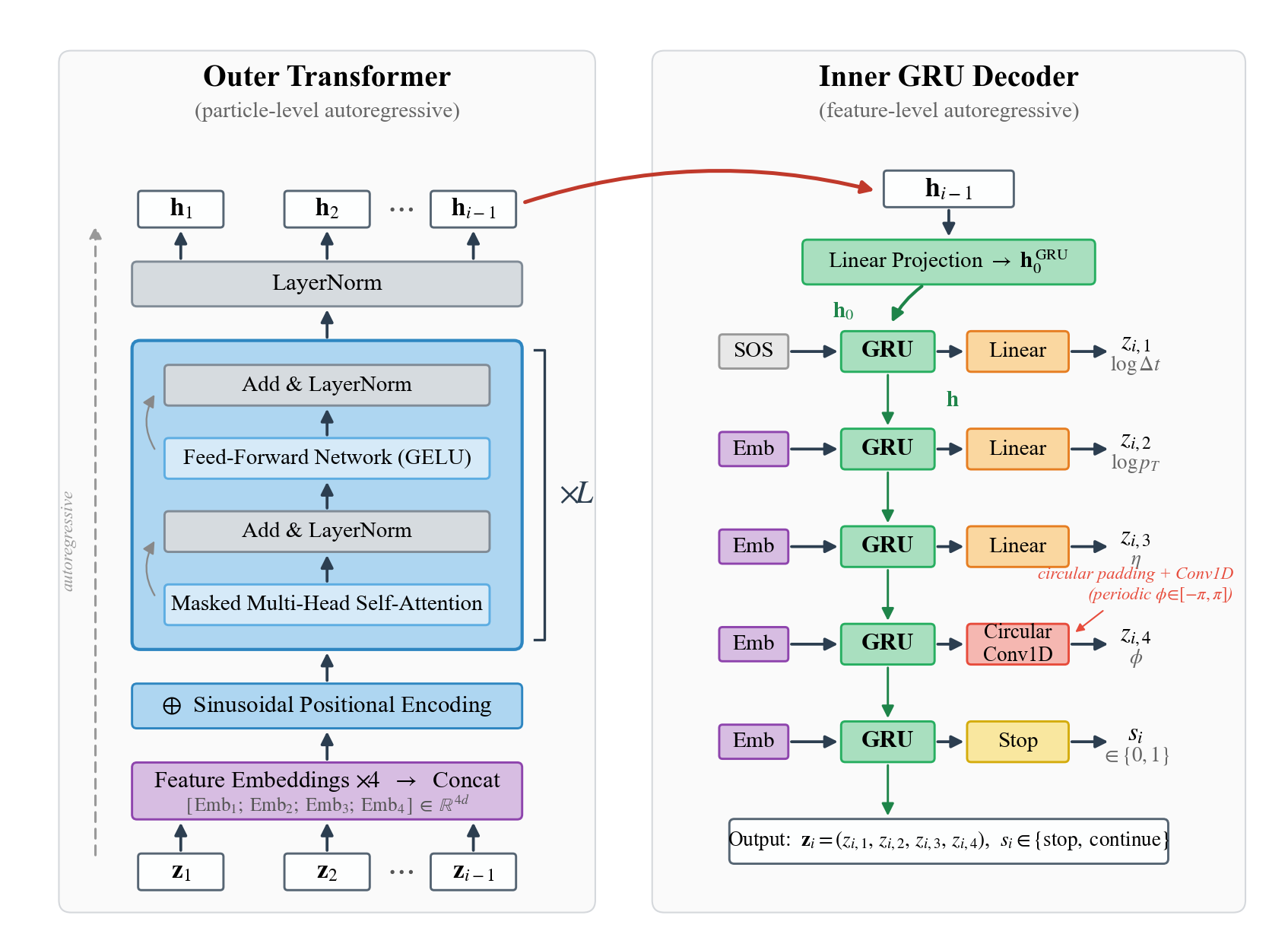}
		\caption{Architecture of the Nested-GPT model.
			\emph{Left:} the outer Transformer operates at the particle level. Each
			particle token $\mathbf{z}_i$ is embedded via four feature-specific embedding
			layers whose outputs are concatenated into $\mathbb{R}^{4d}$, augmented with
			sinusoidal positional encoding, and processed by $L$ causal Transformer blocks
			(masked multi-head self-attention, feed-forward network, residual connections
			and layer normalization) to produce context vectors
			$\mathbf{h}_1,\dots,\mathbf{h}_{i-1}$.
			\emph{Right:} the inner GRU decoder generates the features of particle $i$
			sequentially, conditioned on the context $\mathbf{h}_{i-1}$ via a learned
			linear projection that initializes the GRU hidden state
			$\mathbf{h}_0^{\mathrm{GRU}}$. At each step the GRU produces logits through
			linear output heads for the four discretized features. After all four features are generated, a final
			stop head outputs the stop/continue decision $s_i\in\{0,1\}$.}
		\label{fig:nested-gpt-arch}
	\end{figure*}

	Our model is hierarchical: an outer autoregressive Transformer models particle-to-particle dependencies, while an inner autoregressive decoder models the feature-by-feature structure within a particle. Conditioned on the previously generated particles $\mathbf{z}_{<i}$, the model predicts a distribution over the next particle and an explicit stop/continue decision $s_i\in\{0,1\}$,
	\begin{align}
		p_{\theta}\left(s_i,\mathbf{z}_{i}|\mathbf{z}_{<i}\right)=p_{\theta}\left(s_{i}|\mathbf{z}_{\leq i}\right)\prod_{f=1}^{F}p_{\theta}\left(z_{i,f}|\bm{z}_{i,<f},\mathbf{z}_{<i}\right)
		\label{eq:gpt-factorization}
	\end{align}
	The stop token enables variable-multiplicity event generation; sampling terminates once $s_i=0$ is produced and $\bm{z}_{i,<j}=(z_{i,1},\dots,z_{i,j-1},0,\dots)$. The full architecture is shown in Fig.~\ref{fig:nested-gpt-arch}.
	
	As shown in the left panel of Fig.~\ref{fig:nested-gpt-arch}, the outer model is
	a GPT-like Transformer decoder with causal self-attention. Each particle token
	vector
	$\mathbf{z}_i$ is embedded by concatenating feature-specific embeddings,
	\begin{equation}
		\mathbf{h}^{(0)}_i
		= \bigl[\mathrm{Emb}_1(z_{i,1});\,\dots;\,\mathrm{Emb}_F(z_{i,F})\bigr]
		\in\mathbb{R}^{F d},
		\label{eq:gpt-embed}
	\end{equation}
	followed by a sinusoidal positional encoding in the particle index $i$.
	Stacked Transformer blocks---each consisting of masked multi-head self-attention,
	a feed-forward network with GELU activation,
	and residual connections with layer normalization---produce
	context vectors $\mathbf{h}_i$ such that $\mathbf{h}_i$
	depends only on $\mathbf{z}_{\le i}$ via a triangular causal mask
	\cite{Vaswani:2017attention,Hendrycks:2016gelu,Ba:2016layernorm}. In this work
	we use $N_{\max}=100$, $B=256$ bins per feature, embedding dimension $d=64$, and
	$6$ Transformer layers with $8$ attention heads (dropout $0.3$).
	
	The inner decoder (right panel of Fig.~\ref{fig:nested-gpt-arch}) generates the
	four
	features of the next particle sequentially. The context is first mapped to the
	initial Gated Recurrent Unit (GRU) hidden state $\mathbf{h}_0^{\mathrm{GRU}}$ through a learned linear
	projection. A $2$-layer GRU with hidden size
	$256$ then steps through the
	features in the order $(\log\Delta t,\,\log p_T,\,\eta,\,\phi)$: at each step
	the GRU takes as input an embedding of the previously generated feature (or a
	start-of-sequence token for the first feature) and outputs logits via linear
	heads \cite{Cho:2014emnlp,Chung:2014gru}. After all four features are generated, a final output
	head produces the stop/continue decision $s_i\in\{0,1\}$.
	
	We train the model with teacher forcing, predicting particle $i$ given the
	ground-truth prefix $\mathbf{z}_{<i}$. The primary term is a categorical
	cross-entropy over bins. Since the bins are ordered, we use an ordinal-aware target distribution
	by smoothing the one-hot label with a Gaussian kernel
	in bin index space \cite{Diaz:2019softlabels}. For a target bin $z$ we set
	\begin{align}
		\tilde{y}_{f,k}(z)
		&= \frac{\exp\!\left[-\tfrac{1}{2}\bigl(\Delta_{f}(k,z)/\sigma_f\bigr)^2\right]}
		{\sum_{\ell=0}^{B-1}\exp\!\left[-\tfrac{1}{2}\bigl(\Delta_{f}(\ell,z)/\sigma_f\bigr)^2\right]},
		\label{eq:gpt-softlabels} \\
		\Delta_{f}(k,z) &= k - z,
	\end{align}
	where $\sigma_f$ sets the smoothing scale.
	
	Let $p_{\theta,f,k}$ be the softmax probability for feature $f$ and bin $k$, and
	let $m_i\in\{0,1\}$ be the particle mask. The ordinal cross-entropy term is
	\begin{equation}
		\mathcal{L}_{\mathrm{ord}}
		= -\frac{1}{F\sum_i m_i}\sum_i\sum_{f=1}^{F} m_i \sum_{k=0}^{B-1} \tilde{y}_{f,k}(z_{i,f}) \log p_{\theta,f,k}.
		\label{eq:gpt-ordloss}
	\end{equation}
	In addition, we include a regression loss on bin centers by matching the
	predicted mean $\hat{q}_{i,f}=\sum_k p_{\theta,f,k}\,c_{f,k}$ to the target
	center $c_{f,z_{i,f}}$, and a standard cross-entropy loss $\mathcal{L}_{\mathrm{stop}}$ for the stop
	token. The total loss is
	\begin{equation}
		\mathcal{L}_{\mathrm{GPT}} = \mathcal{L}_{\mathrm{ord}} + \lambda_{\mathrm{reg}}\mathcal{L}_{\mathrm{reg}} + \lambda_{\mathrm{stop}}\mathcal{L}_{\mathrm{stop}},
		\label{eq:gpt-totalloss}
	\end{equation}
	with default weights $\lambda_{\mathrm{reg}}=0.1$ and $\lambda_{\mathrm{stop}}=0.2$.
	
	While the loss in Eq.~\eqref{eq:gpt-totalloss} trains the model to capture
	inter-particle correlations, the autoregressive factorization in
	Eq.~\eqref{eq:gpt-factorization} conditions each particle on its predecessors
	and therefore does not provide a learned prior for the marginal distribution of
	the leading emission. We therefore introduce a lightweight auxiliary
	autoregressive model for the leading particle and use it to initialize
	generation. Using the same discretized feature representation as above, this
	model factorizes the joint distribution as
	\begin{align}
		p_{\psi}\left(\mathbf{z}_{1},s_{1}\right)=p_{\psi}\left(s_{1}|\mathbf{z}_{1}\right)\prod p_{\psi}\left(z_{1,i}|\bm{z}_{1,<i}\right),
		\label{eq:first-particle-factorization}
	\end{align}
	where $\mathbf{z}_1$ denotes the four discretized leading-emission features and
	$s_1$ the corresponding stop/continue label. In practice we implement
	$p_\psi$ as a small GRU-based autoregressive network trained on the extracted
	first-emission sample with the same feature ordering
	$(\log\Delta t,\log p_T,\eta,\phi)$ and a standard cross-entropy loss for the
	stop label. This auxiliary model provides the learned leading-emission prior
	for the Nested-GPT event generator.
	
	In generation, the leading emission is first sampled from $p_\psi$. If its stop
	label terminates the sequence, the event contains only this emission.
	Otherwise, subsequent particles are generated autoregressively
	(temperature $T=1$) using
	Eq.~\eqref{eq:gpt-factorization} until a stop token is produced or $N_{\max}$ is
	reached; events that reach $N_{\max}$ without stopping are discarded. To recover
	continuous kinematics from the discrete bins, we sample uniformly within each
	bin interval $[e_{f,k},e_{f,k+1}]$ before mapping back to physical observables
	for evaluation.
	
	In the next-particle shift used for training, stop
	labels are aligned with the predicted particle. We use AdamW with a cosine learning-rate schedule and
	mixed-precision training under DDP \cite{Loshchilov:2019adamw}, with
	$(\beta_1,\beta_2)=(0.9,0.95)$ and weight decay $0.01$; the base
	learning rate is $10^{-5}$ and is scaled linearly with the number of GPUs.  The
	learning rate is held constant for the first $25\%$ of epochs and then decayed
	with a cosine schedule to $1\%$ of its initial value. Unless stated otherwise we
	use a Gaussian smoothing scale $\sigma=2.0$, along with loss weights
	$(\lambda_{\mathrm{reg}},\lambda_{\mathrm{stop}})=(0.1,0.2)$. We use the same
	event-level $90\%/10\%$ train/validation split as in the flow-matching baseline.
	
	\section{Results}
	\label{sec:results}
	
	We evaluate the fidelity of the learned event generators by comparing their output with reference samples from \texttt{NGL-resum}~\cite{Balsiger:2020ogy}, which provides a numerical solution to the large-$N_c$ NGL evolution via a Markovian dipole shower. Our validation proceeds in two stages: first, we assess the capacity of Nested-GPT to explicitly encode a phase-space distribution defined by a fixed veto condition. Subsequently, we perform a more stringent generalization test to determine whether the model preserves the intrinsic shower structure when the training environment and the final analysis prescriptions are decoupled.

	\subsection{Events with a fixed gap}
	\label{sec:fixed-gap}
	
	We initially benchmark the model on an event ensemble defined by a fixed veto region, \(|y|<0.8\). Following the \texttt{NGL-resum} algorithm, an event is terminated once an emission enters this central rapidity interval, though the veto-triggering emission itself is retained in the sequence. Nested-GPT is trained directly on this truncated sample to determine whether the architecture can internalize the boundary conditions and termination logic of the shower.

	\begin{figure}[t]
		\centering
		\includegraphics[width=0.85\columnwidth]{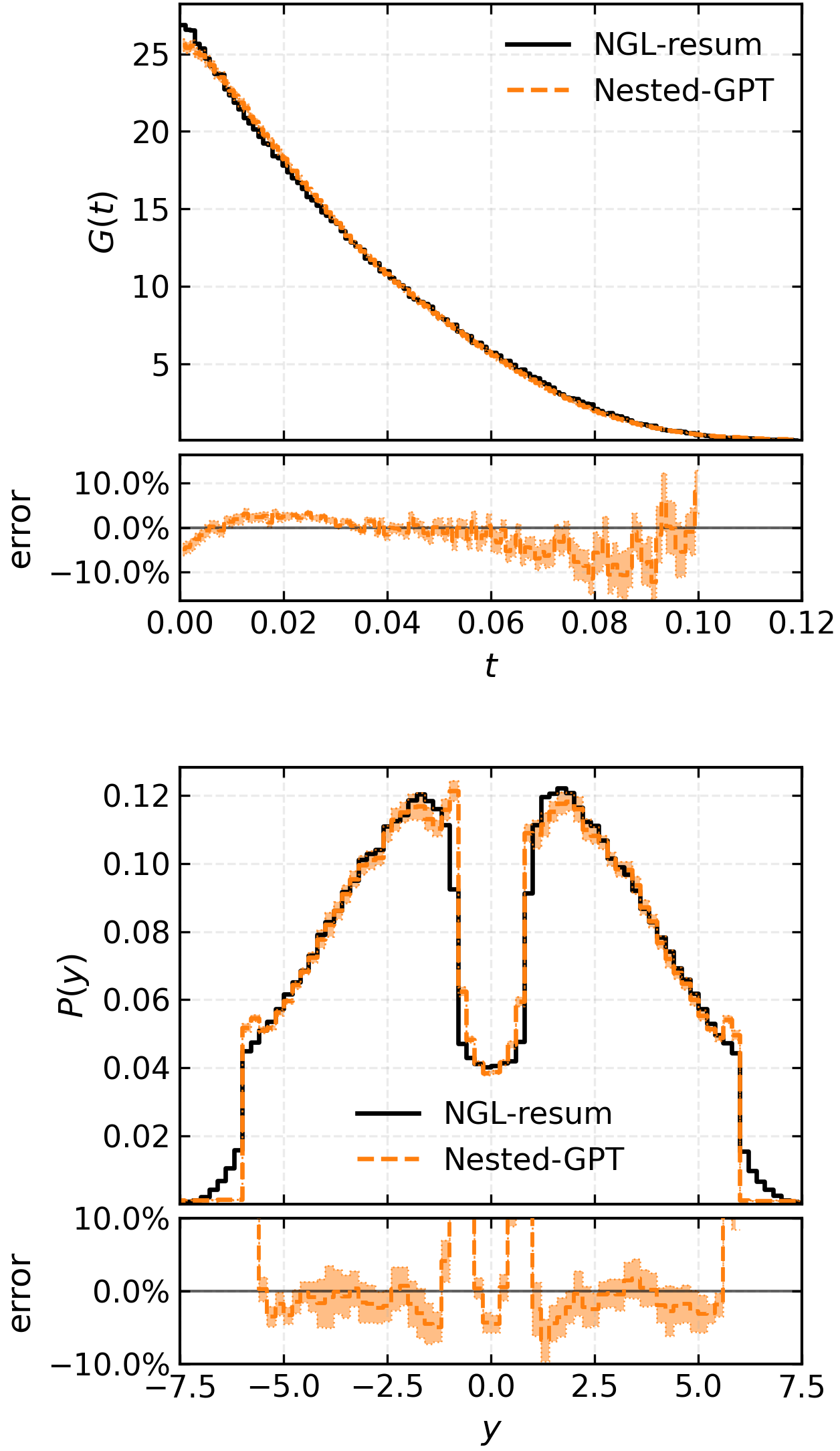}
		\caption{Comparison of the gap fraction $G(t)$ and normalized rapidity density $P(y)$ generated by Nested-GPT against the \texttt{NGL-resum} reference with a veto condition of $|y|<0.8$. Shaded bands indicate the standard deviation across samples generated from 10 consecutive training epochs.}
		\label{fig:truncated-nested-gpt-direct}
	\end{figure}

	In \cref{fig:truncated-nested-gpt-direct}, we compare the gap fraction \(G(t)\) and the normalized rapidity density \(P(y)\) generated by Nested-GPT against the \texttt{NGL-resum} reference. The shaded bands denote the bin-wise standard deviation across $10$ consecutive training checkpoints. This serves as a metric for the numerical stability and convergence of the learned distributions.
	
	The model exhibits excellent agreement across the entire kinematic range, accurately reproducing the evolution of the gap fraction. Marginal statistical fluctuations in \(G(t)\) at large evolution times \(t\) correspond to the deep infrared regime, where the gap fraction is asymptotically suppressed and the available sample size diminishes. For the rapidity distribution \(P(y)\), Nested-GPT accurately captures the characteristic depletion in the central region induced by the veto. Furthermore, the model captures the sharp falloff at large \(|y|\) originating from the $|y|<6$ collinear regulator in the reference sample. Notably, the non-vanishing density at \(y \approx 0\) is correctly recovered, reflecting the inclusion of the boundary-crossing emission as dictated by the veto-terminated dipole-shower logic~\cite{Balsiger:2018ezi, Balsiger:2020ogy}. This confirms that the hierarchical autoregressive structure of Nested-GPT can successfully internalize the complex phase-space restrictions and termination conditions inherent in non-global evolution.
	
\begin{figure}[t]
		\centering
		\includegraphics[width=0.85\columnwidth]{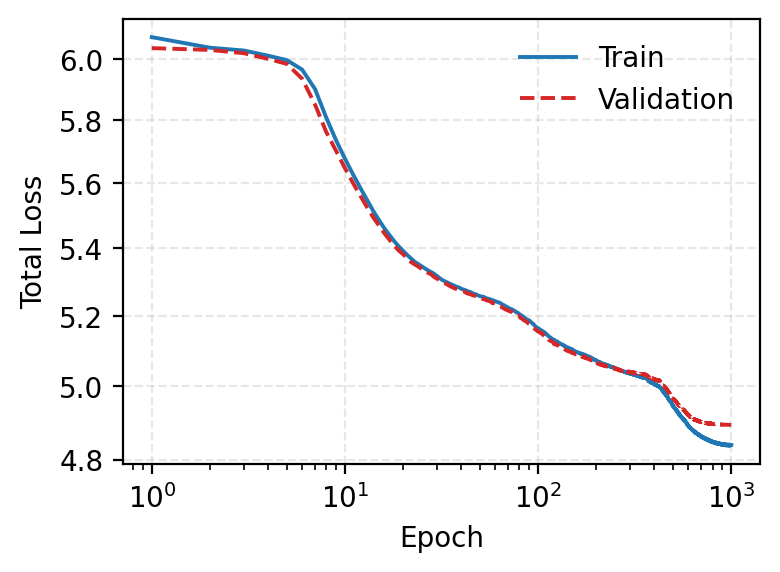}
		\caption{Training history of the Nested-GPT model on the fixed-gap dataset. The validation loss begins to diverge from the training loss near epoch 1000, indicating the onset of overfitting.}
		\label{fig:truncated-loss-curve}
	\end{figure}

	The corresponding training trajectories, illustrated in \cref{fig:truncated-loss-curve}, exhibit a punctuated convergence profile characterized by distinct plateaus followed by rapid descents. This stepwise reduction suggests a hierarchical learning process, wherein the model progressively resolves complex correlation patterns within the ordered emission sequence. Beyond epoch 1000, a visible divergence between the training and validation losses emerges. This separation, a classic manifestation of nascent overfitting, marks the threshold beyond which further training yields diminishing returns for generalization.
	
	\subsection{Generalization}
	
	\begin{figure}[t]
		\centering
		\includegraphics[width=0.85\columnwidth]{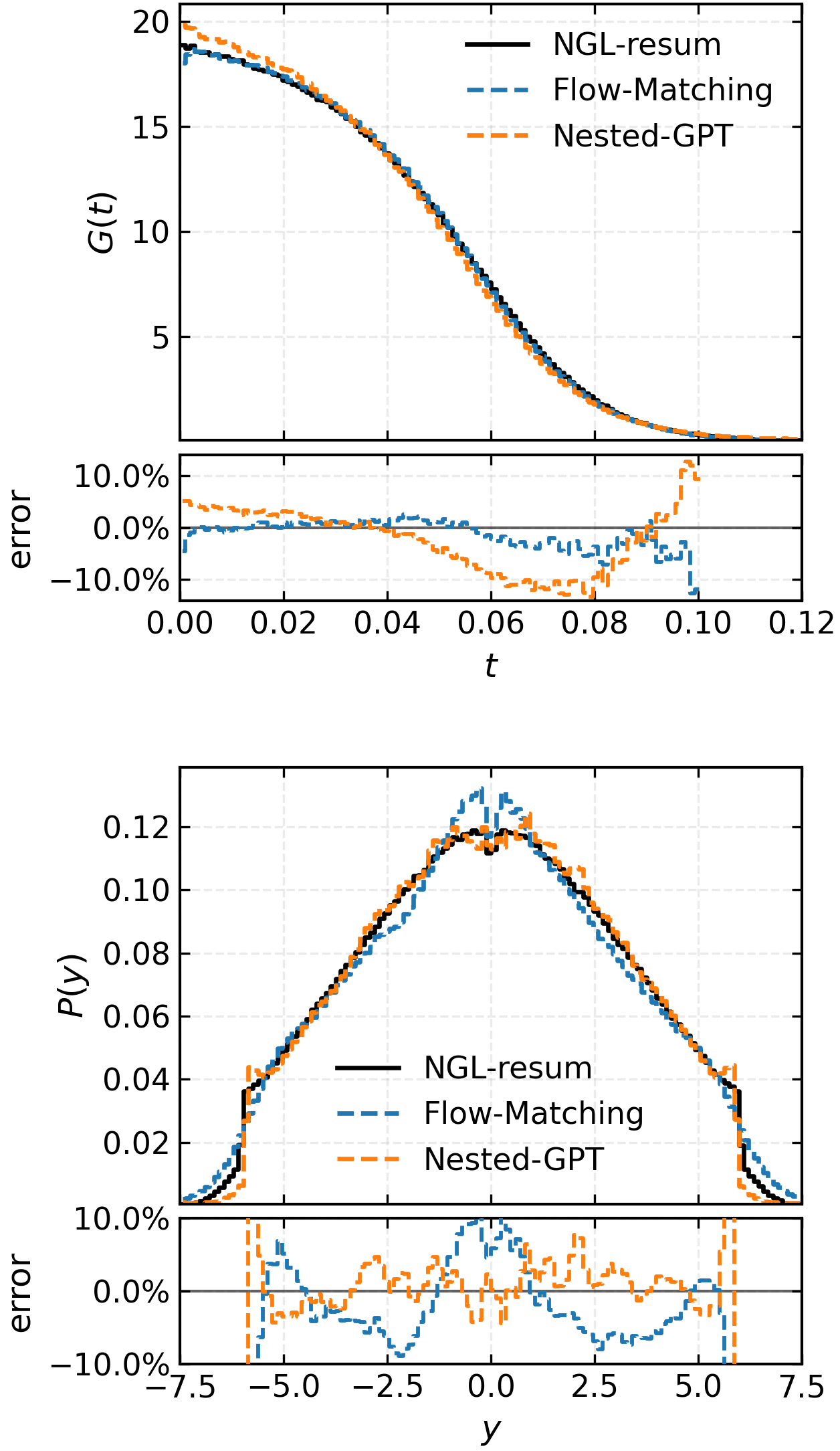}
		\caption{Comparison of the inclusive shower samples generated by the flow-matching baseline and the Nested-GPT model. Both networks are trained on dipole-shower events with a negligible veto constraint and benchmarked against the \texttt{NGL-resum} reference.}
		\label{fig:full-sample-comparison}
	\end{figure}
	
	A robust AI-driven event generator must transcend the direct memorization of its specific training domain. To rigorously validate the predictive power of our approach, we investigate whether the learned architectures can accurately adapt to phase-space constraints that were explicitly absent during training. We train the models on an inclusive event configuration to test whether they have sufficiently modeled the fundamental Markovian branching dynamics to accurately predict highly exclusive, vetoed observables post-generation.
	
	We lift the gap veto condition from the training sample, utilizing instead an inclusive sample generated with essentially no veto constraint. \footnote{Because a strictly vanishing veto leads to an unphysical configuration in NGL evolution, a negligible algorithmic cutoff of \( |y|<0.01 \) is applied during generation.} In this generalization test, we evaluate both the Nested-GPT architecture and the flow-matching baseline. We first verify baseline fidelity by examining the unrestricted training space itself. As demonstrated in \cref{fig:full-sample-comparison}, both models exhibit excellent agreement with the \texttt{NGL-resum} reference distributions, accurately capturing the inclusive gap fraction \(G(t)\) and rapidity density \(P(y)\).
	
	\begin{figure}[t]
		\centering
		\includegraphics[width=0.85\columnwidth]{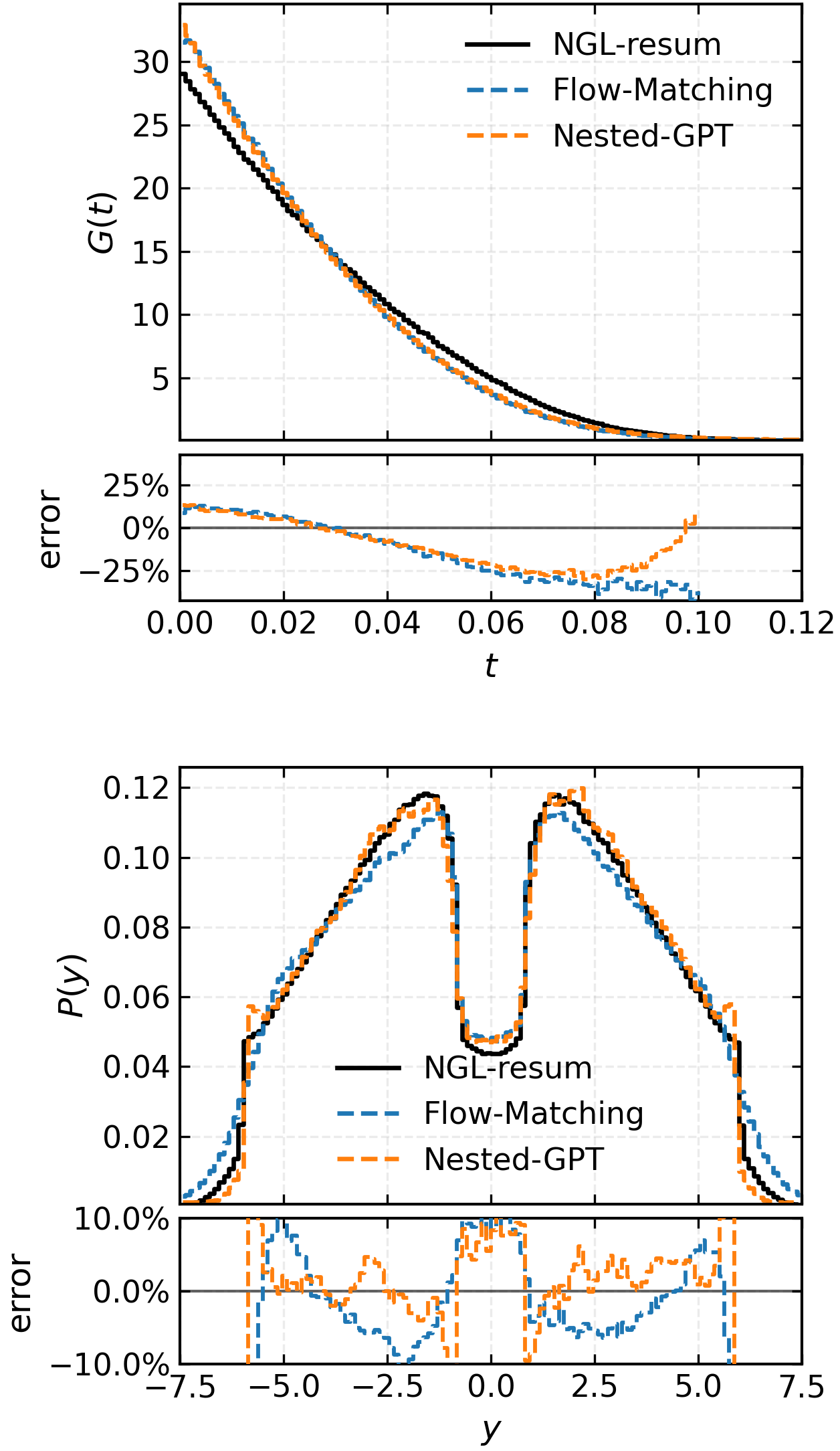}
		\caption{Events from both models are terminated after generation at the first emission entering the central region, \( |y|<0.8 \). The resulting distributions are compared with the \texttt{NGL-resum} events subjected to the same prescription.}
		\label{fig:truncation-test}
	\end{figure}
	
	We then impose a post-hoc analysis prescription identical to the fixed-gap scenario in \cref{sec:fixed-gap}: generation is unconstrained, but the predicted event is truncated offline at the first emission entering the veto region (\( |y|<0.8 \)). The resulting distributions are compared with \texttt{NGL-resum} events subjected to the identical offline cut in \cref{fig:truncation-test}.
	
	Remarkably, both models successfully recover the main features of the vetoed sample. The rapidity density \(P(y)\) develops the characteristic depletion around \(y\simeq 0\), alongside the associated symmetric two-peak structure. Similarly, the gap fraction \(G(t)\) exhibits robust agreement over the populated kinematic regime. Compared with the direct training results in \cref{fig:truncated-nested-gpt-direct}, the overall fidelity of both architectures shows no significant deterioration. 
	
	This outcome is nontrivial because correctly imposing the post-hoc veto condition requires identifying the \emph{chronologically first} emission to enter the gap. Thus, accurately reproducing the truncated macroscopic distributions requires the models to have captured the causal, time-ordered shower sequence, rather than merely fitting inclusive single-emission probability densities.
	The empirical fact that both architectures can reconstruct these exclusive distributions after being trained on an inclusive sample demonstrates that the essential chronological ordering information is robustly preserved. This confirms that both the established continuous flow-matching paradigm and the novel hierarchical autoregressive approach exhibit the ability to generalize from inclusive shower histories to vetoed observables whose phase-space restrictions were absent during training.
	
	\section{Conclusion}
	\label{sec:conclusion}
	
	In this paper, we have explored the integration of generative AI architectures with the resummation of NGLs via large-$N_c$ dipole showers. At LL accuracy and in the large-$N_c$ limit, the non-global RG evolution admits a stochastic dipole-shower realization for time-ordered soft-emission sequences. Taking this as our theoretical foundation, we developed Nested-GPT, a hierarchical autoregressive framework for variable-multiplicity dipole-shower histories, and benchmarked it against an established Transformer flow-matching baseline.
	
	Our results demonstrate that both modeling strategies can reproduce the emission-level correlations of the reference dipole shower for the observables studied. The flow-matching framework proves to be a robust benchmark for continuous feature learning, while the Nested-GPT model introduces an alternative, hierarchical autoregressive approach that dynamically manages event termination. The results show that both architectures achieve comparable levels of physical fidelity, accurately capturing the non-linear evolution of gap fractions and rapidity distributions. 
	
	The primary methodological contribution of this work is the demonstration that chronological parton-shower histories can be efficiently emulated through causal attention mechanisms. Nested-GPT maintains a level of performance similar to the established baseline, while its novel architecture provides a new conceptual route for simulating variable-multiplicity stochastic sequences. This flexibility suggests that generative AI offers a versatile toolkit for emulating precise but computationally demanding resummation frameworks. While the present study focused on the LL large-$N_c$ dipole shower as a proof of concept, the most compelling avenue for future research lies in deploying these architectures to bypass the severe combinatorial and computational bottlenecks encountered in finite-$N_c$ amplitude evolution, spin-correlated branchings, and NLL parton showers.
	
	%%%%%%%%%%%%%%%%%%%%%%%%%%%%%%%%%%%%%%%%%%%%%%%%%%%%%%%%%%%%%%%%%%%%%%%%%%%%%%%%
	{\it Acknowledgments --}
	%%%%%%%%%%%%%%%%%%%%%%%%%%%%%%%%%%%%%%%%%%%%%%%%%%%%%%%%%%%%%%%%%%%%%%%%%%%%%%%%
	The authors thank Mei-Sen Gao for his contribution during the early stages of this work, and Tian-Ji Cai for the helpful discussions. This work is supported by the National Natural Science Foundation of China under Grant No.~12275052, No.~12147101, No.~12547102, No.~124B1003. W.L. is also supported by the China Postdoctoral Science Foundation under grant No.~2025M783369. D.Y.S. is also supported by the Innovation Program for Quantum Science and Technology under grant No.~2024ZD0300101.

	\bibliography{refs}

\end{document}